\documentclass[11pt,twoside]{atmp}

\usepackage{amsmath,amssymb}
\usepackage[all]{xy}
\usepackage{color}
\usepackage{graphicx}

\begin{document}

\title[Cyclotron braid group]
{Cyclotron braid group approach to Laughlin correlations}


\author[J. Jacak, I. J\'o\'zwiak, L. Jacak, K. Wieczorek]{J. Jacak$^1$, I. J\'o\'zwiak$^2$, L. Jacak$^{1*}$, K. Wieczorek$^1$}

\address{$^1$ Institute of Physics, $^2$ Institute of Computer Science, \\
Wroc{\l}aw University of Technology,\\
Wyb. Wyspia\'nskiego 27, 50-370 Wroc{\l}aw, Poland}
\addressemail{$^{*}$ ljacak@pwr.wroc.pl}

\begin{abstract}
Homotopy braid group description including  cyclotron motion of char\-ged interacting
 2D particles at strong magnetic field presence  is developed in order to explain, in algebraic topology terms,
Laughlin correlations in fractional quantum  Hall systems.
There are introduced special cyclotron braid subgroups of a full braid group with
 one dimensional unitary representations suitable to satisfy Laughlin correlation requirements.
In this way an implementation
       of composite fermions (fermions with auxiliary  flux quanta attached in order to
         reproduce Laughlin correlations)  is formulated within uniform for all 2D particles braid group approach. The
fictitious fluxes---vortices attached to the composite fermions in a traditional formulation are replaced with additional cyclotron trajectory loops
unavoidably occurring when ordinary cyclotron radius is too short in comparison to particle separation and does not allow for particle interchanges
along single-loop cyclotron braids. Additional loops enhance the effective cyclotron radius and restore particle interchanges.
A new type of 2D particles---composite anyons is also defined via unitary representations of cyclotron
braid subgroups.
It is demonstrated that composite fermions and composite anyons are rightful 2D particles, not  auxiliary compositions
with fictitious fluxes and are associated with  cyclotron braid
subgroups instead of the full braid group,
which may open also a new opportunity for non-Abelian composite anyons for topological quantum information processing applications, due to richer
representations of subgroup than of a group.

\end{abstract}

\maketitle



\section{Introduction}

Specific topological properties of 2D $N$-particle  systems have been recognized
within algebraic topology \cite{spanier} using homotopy group methods \cite{birman,mermin}. They turned out to be
 of particular significance in understanding of quantum behavior of 2D electron systems widely experimentally
 investigated in Hall configuration  since the discovery of fractional quantum Hall effect (FQHE) \cite{tsui}.
A hierarchy of Landau level (LL) fractional fillings was
observed \cite{tsui,pin,prange} and explained by new topological concepts closely related with planar geometry \cite{laughlin1,laughlin2,wilczek}.
The exceptional topology of 2D systems is connected with nontrivial homotopy
groups \cite{birman,artin,jac1,imbo,enar} describing  planar trajectories for many-particle systems. Classes of  topologically nonequivalent closed loops in the
configuration space of a system of $N$ identical  particles build up  $\pi_1$ homotopy group \cite{spanier},
called in this case a braid group (full and pure for
indistinguishable and distinguishable particles, respectively) \cite{birman,artin,jac1,imbo}.
Braid groups for 2D case are infinite, while for higher dimensions are finite and  equal (full braid groups) to the permutation group $S_N$
(for 1D case
this formalism is irrelevant) \cite{mermin}. Unitary representations, in particular one-dimensional  (1DURs) of the full braid group serve for identification
of quantum particles corresponding to the same classical ones \cite{wilczek}. For
$S_N$ there exist only two distinct 1DURs: $\sigma_i\rightarrow e^{i\pi}$ or
 $\sigma_i\rightarrow e^{i0}$, ($\sigma_i$---the  interchange of $i$th and $(i+1)$th particles)
 corresponding to fermions and bosons, respectively.
In 2D there is, however, an abundance of distinct 1DURs of the full braid groups: $e^{i\theta}$, $\theta\in(-\pi, \pi]$, associated
with anyons (Abelian) \cite{laughlin1,laughlin2,pin,wilczek,dassar}. Anyons reveal  a
fractional statistics---the interchange of two anyons results in a $\theta$ phase shift of the wave function \cite{wilczek,wu}.

Crucial for understanding
of FQHE was  the formulation by Laughlin \cite{laughlin1,laughlin2} of the wave function for a ground state of 2D charged particle
system at strong magnetic field presence.
The Laughlin function \cite{laughlin2}  corresponds to $\frac{1}{p}$ ($p$-odd integer)  fractional filling  of the lowest LL
and is a generalization of the Slater determinant. The Slater function for completely filled lowest LL, for magnetic field in
cylindrical gauge, has   the form (up to an exponential factor) of the Vandermonde determinant,   $\prod\limits_{i<j}(z_i-z_j)$,
$z_i=x_i+iy_i$ stands here for $i$th 2D particle coordinate  expressed as
a complex number.
Replacement in this Slater function of the Vandermonde determinant with
the Jastrow polynomial, $\prod\limits_{i<j}(z_i-z_j)^p$, ($p$ odd integer),
results in the Laughlin wave function \cite{laughlin1,laughlin2} for filling $\frac{1}{p}$ . The Laughlin function is
still antisymmetric  but  differs from the Slater function in the phase shift acquired due to
interchange of a 2D particle pair. For the Vandermonde function it is $\pi$, while for the Jastrow function $p\pi$.
The difference
in phases is important in planar geometry (in higher dimensions the phase shift
has no  meaning), but $2\pi$ periodicity of the phase factor, $e^{ip\pi}=e^{i\pi}$, also in 2D seemingly does not allow for distinguishing
of the statistics imposed by Laughlin correlations from ordinary fermion antisymmetry. Therefore more subtle topological
attitude---the braid group methods, should be here applied in order to grasp the novelty introduced by the Laughlin function.

The phenomenological approach to Laughlin correlations was introduced in terms of
composite fermions (CFs), regarded as ordinary 2D fermions with associated to each particle even number of magnetic flux quanta \cite{jain,hon}.
The even number, $q$, of magnetic flux quanta attached to individual particles does not change antisymmetry of the total system wave function,
but due to Aharonov-Bohm effect results in additional $q\pi$ phase shift during particle pair interchange \cite{wilczek}. In this manner the magnetic field
 local fluxes, called as vortices,
attached to CFs
model the Laughlin correlations \cite{laughlin1,laughlin2}.
The CF attitude suffers, however, from an artificial character of the construction, i.e., not explained  source of the magnetic field fluxes changing
fermions into CFs.

Nevertheless, CFs regarded as only weakly (residually) interacting,
surprisingly well describe Hall systems \cite{jain,hon} especially within
the lowest LL (for higher LLs the inter-level mixing effects perturb a CF picture).
  The vortices of CFs,
 oriented oppositely to the external field are assumed to be able to screen the external magnetic field, and in the effective weaker resultant field,
one can deal
with an integer quantum Hall effect---which yields the fractional hierarchy $\frac{p}{2p\pm 1}$ \cite{jain,hon}. In other words, the oscillations
in Hall conductivity (FQHE) can be  associated with Szubnikov-de Haas oscillations in an effective reduced magnetic field.
    The interesting observation supporting this model is the so-called Hall metal state \cite{halperin} at filling fraction $\frac{1}{2}$,
    when the total external magnetic field should be canceled by the averaged internal field of CF fluxes.
    It still  arises, however, an important question of what is the physical source of these magnetic flux quanta, i.e., vortices, attached to charged particles which
    alter original fermions  into CFs and how to understand localization of magnetic field fluxes on individual particles.

In the present paper we  demonstrate  the braid structure of composite fermions, as particles with statistical properties required by
 Laughlin correlations, via association  them with cyclotron
braid subgroups instead of the full braid groups.
Introduced below cyclotron braid subgroups reflect the classical braid picture for 2D  $N$-particle charged system at the presence
of magnetic field. The quantization, via 1DURs of these cyclotron subgroups, allow for natural explanation of Laughlin correlations,
without invoking  artificial vortices. In particular this approach elucidates
the CF construction and the true character of auxiliary Jain's vortices \cite{jain}, which turned out  a useful model of
basic trajectory loops unavoidably occurring on cyclotron braids at fractional LL fillings $\frac{1}{p}$. The multi-loop braids from cyclotron
braid subgroups  allow for particle interchange in the braid picture,  when the single-loop cyclotron diameter
              is shorter than the particle separation, which precludes their exchanges along single-loop cyclotron
              trajectories.
               In order to enhance cyclotron radius and to restore particle interchanges in braid picture, each particle must
             traverse, in classical braid meaning, a closed $p$-loop cyclotron trajectory, or in quantum language,
              each particle takes away $p$   quanta of the external magnetic field flux; $p-1$ of them
            play the equivalent role as  $p-1$ flux quanta attached to each CF in a traditional model, reducing the external field.
Topological implementation of CFs in braid group terms was not previously formulated due to periodicity of 1DURs.
 Association of composite particles (including composite fermions) with a separate cyclotron braid subgroups
 allows, however,  for distinguishing them in terms of unitary representations, despite $2\pi$ periodicity of the unitary factor.

The  paper is organized as follows. In the next paragraph the main lines of the braid group approach to quantum systems
are summarized. In the following one, the original idea of cyclotron braid subgroups is developed
and applied to description of CFs, and more generally to composite-anyons. The multi-loop structure
of cyclotron braids, essential for CF description, is explained.  The role of the Coulomb interaction is described in a separate paragraph.
The possible application of introduced composite anyons to topological quantum information processing (QIP) is indicated.

 \section{Braid group method for description of $N$-particle systems}

               \subsection*{Definitions of a full and a pure braid groups}

               Braid group is a first homotopy group \cite{spanier}, $\pi_1$, for configuration space of $N$-particle system. $\pi_1(A)$ is a group
               of topologically nonequivalent classes of closed trajectories in the space A. In the case  of   $N$-particle system, $A$
               is an appropriate classical configuration space. The braid groups display only
               a possible classical motion of $N$-particle system and a quantization
               is  performed via unitary representations of classical braid trajectories, as it is described below.

The configuration space  of $N$  identical particles located on a manifold $M$ (e.g., $R^n$, or compact manifolds) is defined as:
$Q_N(M)=(M^N\setminus\Delta)/S_N$,
 for indistinguishable  identical particles, and as:
 $F_N(M)=M^N\setminus\Delta$,
 for distinguishable  identical particles;
$M^N$ is the $N$th Cartesian product of the manifold $M$, $\Delta$  is the set of diagonal points
 (when coordinates of two or more particles coincide), subtracted in order to preserve conservation of the particle number,
 $S_N$   is the permutation group---the quotient structure  is introduced in order to account for indistinguishability
 of quantum particles.
 Note, that indistinguishability of particles is here artificially introduced in the definition of configuration space, which indicates that this
 property is independent of quantum uncertainty principles.

  For these configuration spaces two types of braid groups are defined \cite{birman}:
\begin{equation}
\pi_1(Q_N(M))=\pi_1(M^N\setminus\Delta)/S_N),
\end{equation}
a full  braid group and
\begin{equation}
\pi_1(F_N(M))=\pi_1(M^N\setminus\Delta ),
\end{equation}
 a pure braid group.

 For $M=R^n$, $n>2$ the braid group have a simple structure. The full braid group, for $n>2$,  equals to a permutation group $S_N$ (note, that this group is a
  finite group,
 of rank $N!$).
 For $M=R^2$ (and for compact locally 2D manifolds, as a sphere or a torus in 3D) the braid groups are infinite highly nontrivial groups.

\begin{figure}[tb]
\unitlength 1mm
\begin{center}
\begin{picture}(120,23)
\put(0,0){\resizebox{120mm}{!}{\includegraphics{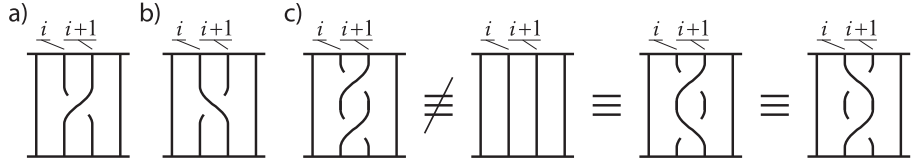}}}
\end{picture}
\end{center}
\caption{\label{fig1} Geometrical braid presentation for $B_N$: the generator $\sigma_i$ (a) and its inverse, $\sigma_i^{-1}$ (b);
square of the generator $\sigma_i^2$ (c)}
\end{figure}

	It is convenient to illustrate a structure of the braid groups for the plane via a simple presentation using
	 geometrical braids \cite{birman,jac1}---cf.  Fig. \ref{fig1}.
	In this figure there are depicted: (a)
	geometrical braid  corresponding to the generator $\sigma_i$ of the full braid group  (interchange of the  $i$th and  $(i+1)$th strings representing
	particle trajectories),  (b) geometrical braid corresponding to  the inverse element
	of the generator, $\sigma_i^{-1}$, (c) geometrical braid for the  square of the generator $(\sigma_i)^2\neq e$ ($e$---the neutral element of the group).
	In 3D $(\sigma_i)^2=e$, which simplifies the braid structure to ordinary permutation group $S_N$, while in 2D  $(\sigma_i)^2\neq e$ and it causes
          complicated (of infinite type) structure of planar braids.

One can list
formal conditions imposed on generators $\sigma_i,\; i=1,...,N-1$, in order to
define the full braid group for the plane, in an abstract manner \cite{artin,birman}. These conditions are written below  and are illustrated by geometrical braids in  Fig. \ref{fig3}
a,b,
	 \begin{equation}
	 \label{e10a}
	 \begin{array}{ll}
	 \sigma_i\sigma_{i+1}\sigma_i=\sigma_{i+1}\sigma_{i}\sigma_{i+1},\; & {\rm for}\; 1\leq i\leq N-2,\;\;\;\;\;\\
	 \end{array}
	 \end{equation}
     \begin{equation}
     \label{e10b}
     \begin{array}{ll}
     \sigma_i\sigma_j= \sigma_j\sigma_i.\; & {\rm for}\;1\leq i,j\leq N-1,\; |i-j|\geq 2.\\
	 \end{array}
	 \end{equation}

The initial ordering of particles is not conserved for braids from a full braid group, while for braids from a pure one, the ordering must be conserved.
The generators $l_{ij}$  of a pure braid group \cite{birman} correspond to double exchanges of particle pairs, $ij$,
however,  without any perturbation of the assumed ordering of particles, and have the following form in terms of $\sigma_i$ generators:
	 \begin{equation}
	 l_{ij}= \sigma_{j-1} \cdot \sigma_{j-2}....\sigma_{i+1}\cdot \sigma_i^2\cdot
	 \sigma_{i+1}^{-1}...\sigma_{j-2}^{-1}\cdot \sigma_{j-1}^{-1},\;\; 1\geq i\geq j\geq N-1.
	 \end{equation}
	 The pure group is  a subgroup of the full group since the generators $l_{ij}$
	 are expressed by means of $\sigma_i$ generators.
	 For defining relations for  generators of the pure group cf.  Refs \cite{birman,jac1}.

  Note that the connection between the full braid group and the pure one is given by the quotient
  relation \cite{birman}, $ B_{N}/\pi_1(F_N(R^2))=S_N$ ($B_N$  stands here for commonly used notation for the full braid group for the plane) \cite{artin}.

  For the sphere $S^2$ the additional condition for generators, beyond those given by Eqs \eqref{e10a} and \eqref{e10b}, is imposed \cite{birman},
  \begin{equation}
  \label{sfera}
  \sigma_1\cdot   \sigma_2\cdot... \cdot   \sigma_{n-2}\cdot      \sigma_{n-1}^{2}\cdot        \sigma_{n-2}\cdot ...\cdot      \sigma_2\cdot
    \sigma_1=e,
    \end{equation}
    which displays the fact that on the sphere a loop of a selected particle embracing all other particles is
    contractible to a point. For the torus $T$ additional relations \cite{enar} correspond to two nonequivalent paths
    of each particle on  this not simple-connected manifold.

\subsection*{Quantization in braid group picture}

  Quantization of the system of $N$  identical indistinguishable particles can be performed by
  application of the Feynman integral over trajectories, leading to a propagator (probability for a transition from a point $a$ to a point $b$ in
  the configuration space):
  \begin{equation}
  \label{f11}
  I_{a\rightarrow b}=\int d\lambda e^{iS[\lambda_{a,b}]/\hbar},
  \end{equation}
  where $ S[\lambda_{a,b}]$ is the classical action for the trajectory $\lambda_{a,b}$ in the classical configuration space
  of $N$-particle system, $d\lambda$ is a measure in a trajectory space.
  To each trajectory linking $a$ and $b$ points in the $N$ particle configuration space, one can attach, however, additional
  closed loops which are elements of the full braid group.
  Thus resulting trajectories fall into separated topologically nonequivalent classes, represented by elements of the full braid group.
  Therefore  an
  additional unitary factor (the weight of the separated trajectory class) should be added \cite{wilczek,wu} in the
  formula for integration over trajectories, together with the additional sum over the braid group elements (since each element of the full braid group can be attached
  to a loop-less simple trajectory $\lambda_{a,b}$):
                                               \begin{equation}
                                               \label{f22}
                                               I_{a\rightarrow b}=\sum\limits_{l\in\pi_1} e^{i\alpha_l}\int d\lambda_l e^{iS[\lambda_{l(a,b)}]/\hbar},
                                               \end{equation}
                                               $\pi_1$ represents here the full braid group.
	These factors  $ e^{i\alpha_l}$ form a 1DUR
	 of the full braid group.
	Distinct representations  correspond to distinct types of quantum particles, linked
	to the same classical ones.

As was mentioned in the Introduction, for $S_N$, which is the full braid group for 3D  manifolds (and for higher dimensions), there exist only two distinct
	1DURs,
	\begin{equation}
	\sigma_i \rightarrow \left\{ \begin{array}{l}
	e^{i0},\\
	e^{i\pi},\\
	\end{array}
	\right.
	\end{equation}
	corresponding to bosons and fermions, respectively (leading to a symmetry and antisymmetry properties of relevant wave functions).
For 2D space (the plane), the  braid group (considerably richer than $S_N$)
has an infinite  number of 1DURs \cite{jac1,imbo},  written for the group generators as
$\sigma_i\rightarrow e^{i\theta}, \;\;\theta\in (-\pi,\pi]$,  where
each $\theta$ enumerates a different type of so-called anyons \cite{laughlin1,laughlin2,pin,wilczek,wu,dassar}.  Note that elements of  1DUR   of the full braid group
do not depend on the index $i$ (of the generator $\sigma_i$) owing to the condition \eqref{e10a} imposed on generators. Because the 1DUR elements commute, then
from  Eq. \eqref{e10a}
it follows that $e^{i\theta_i}=e^{i\theta_{i+1}}$, where $\sigma_i\rightarrow e^{i\theta_i}$, which gives this $i$-index independence of 1DUR elements.

For the sphere $S^2$ 1DURs have the form \cite{jac1,imbo},
$e^{i\theta}$, where $\theta=k\pi/(N-1),\;\; k=0,1,2,...,2N-3$. It is interesting to notice, that for two particles on the sphere (i.e., for $N=2$ one has
only $k=0,1$)
only bosonic or fermionic statistics are available (actually because of Eq. \eqref{sfera}), and  anyons may occur on the sphere for
 three particles, at least.
 In the case of a torus $T$, for an arbitrary number $N$ of particles,   $\theta=0$ or $\pi $ are admitted only \cite{imbo,enar}---thus
on a torus any anyons do not exist, except for fermions and bosons. This result was generalized \cite{imbo}
also for all compact locally-2D manifolds with exception for the sphere.

The classical trajectories from the full braid group  have no quantum meaning.
Quantum particles do not traverse any braid trajectories since they do not have trajectories at all.
 In agreement with the general rules of quantization \cite{imbo1,sud}, $N$-particle wave function must transform according to 1DUR of
an appropriate element of the braid group when  the particles traverse classically
a closed loop in the configuration space of $N$-particle system corresponding to this braid element.   As braids from the full braid group describe
interchanges of particles, thus corresponding 1DURs display  statistics phase factors.

Note that important are also multidimensional unitary irreducible representations (MDURs) of braid groups. According to an idea of Kitaev \cite{dassar,kitaev},
an arbitrary
unitary evolution of multi-qubit system (e.g., of a double qubit gate for QIP) \cite{dassar,nielsen} can be approximated by a MDUR
(of an appropriate rank) of a full braid group, provided  the sufficient density level of MDURs in the unitary matrix space \cite{dassar}.
MDURs can be linked with degenerated low-energy excitations (quasiparticles/quasiholes,
typically regarded as anyons) above the ground state for some fractional LL fillings. Since elements of MDUR do not commute, as matrices, these degenerate states of anyons are
referred as non-Abelian anyons \cite{dassar}. Unfortunately, the non-Abelian anyons recently investigated in particular low excited states for 5/2 and 12/5
LL filling factors correspond probably to not sufficiently dense MDURs (for non-Abelian anyons in $\frac{5}{2}$ case the
 MDURs are not dense
              enough to approximate needed qubit gates \cite{dassar}, and another considered now state $\frac{12}{5}$  is still disputable \cite{xia}).
             Thus searching for other opportunities for fractional statistics systems with more dense
MDURs associated with non-Abelian anyons is of high significance. In the next section we will introduce
a cyclotron braid subgroups of a full brad group. As subgroups have usually richer representations than a group, thus one can
expect that the cyclotron braid subgroups would be convenient for topology methods for QIP, since the relevant MDURs of cyclotron subgroups would be more dense
in comparison to representations of a full braid group.

\section{Cyclotron braid groups at magnetic field presence}

Let us emphasize that the braid groups described above are constructed in the absence of the magnetic field.
Elements of the full braid group were  all trajectories without any modifications caused by the  magnetic field.
Inclusion of the magnetic field
considerably confines, however, the variety of admitted trajectories. All trajectories must be of cyclotron shape at the presence
of the magnetic field and this property highly modifies the braid group structure. Instead of a full braid group, cyclotron trajectories
form a braid subgroup---a cyclotron subgroup, in particular at $1/p$ fractional LL filling.
It leads to an opportunity for an implementation of  CFs (2D particles at a strong magnetic field presence)  via cyclotron subgroups of the full braid group.
Following this idea, at
magnetic field presence the summation in the Feynman propagator
must be  confined  to the subgroup elements only, i.e., to selected, suitably to cyclotron motion, classes of trajectories instead of
arbitrary elements of the full braid group.  The 1DURs of the cyclotron braid subgroups will thus substitute
the 1DURs of the full braid group in the path integral \eqref{f22}.

\begin{figure}[tb]
\unitlength 1mm
\begin{center}
\begin{picture}(120,35)
\put(0,0){\resizebox{120mm}{!}{\includegraphics{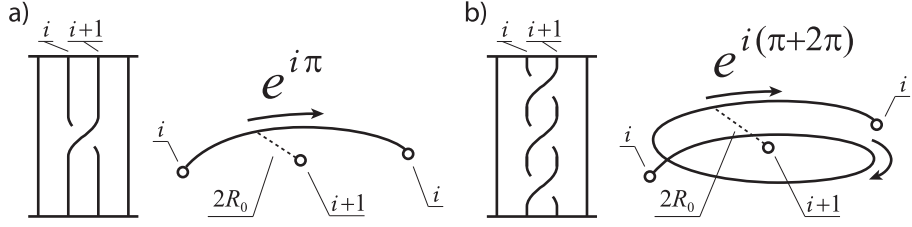}}}
\end{picture}
\end{center}
\caption{\label{fig2}
The generator $\sigma_i$ of the full braid group and the corresponding relative  trajectory of
particle $i$th and $(i+1)$th exchange (a);
 the  generator of the cyclotron braid subgroup, $b_{i}^{(p)}=\sigma_i^p$ (in the figure $p=3$), corresponds to additional $\frac{p-1}{2}$ loops
  when the $i$th particle interchanges with the $(i+1)$th  one (an additional loop results in $2\pi$ phase shift; $2R_0$---inter-particle separation)
 (b)}
\end{figure}

Let us consider 2D charged particle system with planar density $\frac{N}{S}$ ($N$ is the  number of particles, $S$ is the surface of a sample) and at presence of
a perpendicular magnetic field $B$.
Topology of a manifold where the particles are located is assumed here the same as of the plane $R^2$ (it would be considered as an
compact subset of $R^2$, without a  boundary) \cite{birman}.
For this manifold  one can define the full braid group \cite{birman,artin,jac1,imbo}, being the $\pi_1$ homotopy group \cite{spanier}
 of the configuration space
 for $N$ indistinguishable particles
on $R^2$. This braid group is commonly called as $B_N$ (the Artin group) \cite{artin} and is generated by interchanges of neighboring particles at chosen
their ordering \cite{birman,jac1},
\begin{equation}
\sigma_i,\;\; i=1,...,N-1,
\end{equation}
with defining relations given by  Eqs \eqref{e10a} and \eqref{e10b}.

\subsection*{Definition of the cyclotron braid subgroup}

Let us define the cyclotron braid subgroup by means of  its generators
	  $b_i^{(p)}$ of the following form:
	  	 \begin{equation}
	 \label{gen}
	  b_i^{(p)}=\sigma_i^p, \;\;\;p=1,3,5,7,9,...; \;\;\; i=1,...,N-1,
	  \end{equation}
	 where each $p$ corresponds to  a different type
	 of the cyclotron subgroup and $\sigma_i$ are generators of the full braid group.

  The full braid group  element $b_i^{(p)}$ (the generator of the cyclotron braid subgroup of type $p$)   represents the interchange of $i$th and
  $(i+1)$th particles with $\frac{p-1}{2}$ loops. It is clear
	 due to the definition of the single interchanges by the generators $\sigma_i$ of the full braid group, cf. Fig. \ref{fig2}.

\begin{figure}[tb]
\unitlength 1mm
\begin{center}
\begin{picture}(120,98)
\put(0,0){\resizebox{120mm}{!}{\includegraphics{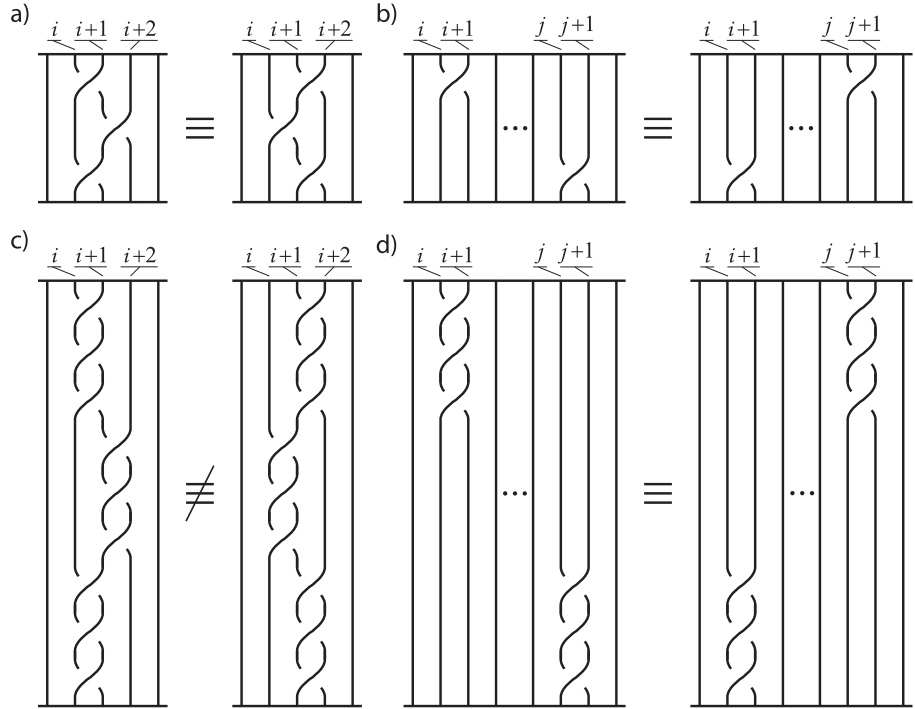}}}
\end{picture}
\end{center}
\caption{\label{fig3}
Formal conditions defining a full braid group for $R^2$, cf. Eqs \eqref{e10a} and \eqref{e10b};
 violation of the condition \eqref{e10a} for the cyclotron subgroup generators $b^{(3)}_i$ (c) (the condition \eqref{e10b} is maintained
 for the cyclotron subgroup generators (d))}
\end{figure}

The  generators $b_i^{(p)}$ create the
	 subgroup of the full braid group as they are expressed by generators $\sigma_i$ of the full braid group.  The $b_i^{(p)}$ do not, however,
	 satisfy the condition \eqref{e10a} (cf. Fig. \ref{fig3} (c)), while the
	 condition \eqref{e10b} is maintained for $b_i^{(p)}$:
	 $  b_i^{(p)}b_j^{(p)}=
	 b_j^{(p)}b_i^{(p)}$, for $1\leq i,j\leq N-1,\;\; |i-j|\geq 2$ (cf.  Fig. \ref{fig3} (d)).

	 The condition \eqref{e10a} resulted in independence of 1DUR of the  braid group generator index $i$.
	 Disappearance of this condition  for the cyclotron braid subgroup leads to
	 possible  dependence of the subgroup 1DUR on the index $i$, in general.
The 1DURs  of the full group, $\sigma_i \rightarrow e^{i\alpha}$, confined to the subgroup, do not depend, however, on $i$ and yield:
\begin{equation}
\label{repr}
	  b_i^{(p)}\rightarrow e^{ip\alpha},
	  \end{equation}
	  $p$ an odd integer,
	  $\alpha \in (-\pi,\pi]$. These 1DURs   of the cyclotron braid subgroup,
	   numbered by the pairs $(p,\;\alpha)$, describe
composite anyons, and, in a particular case, CFs for $\alpha =\pi$.

\subsection*{Multi-loop cyclotron braid  structure}

 For the above construction of the cyclotron subgroups  the $N$-particle wave function acquires an appropriate phase shift
	  due to a peculiar type of particle interchanges in the braid picture, i.e., we replace the Aharonov-Bohm phase of fictitious fluxes
	   by additional braid loops (each loop adds $2\pi$ to the total phase shift---cf. Fig. \ref{fig2}). It is noticeable if
	   one takes into account the rules of quantization in the braid group framework \cite{imbo1,sud}.
	    In agreement with them, $N$-particle wave function must transform according to 1DUR of
an appropriate element of the braid group, when  the particles traverse classically
a closed loop in the configuration space of $N$-particle system corresponding to this braid element.
For  the cyclotron braid subgroup generated by  $b_i^{(p)}$, $i=1,...,N-1$
(defined   by Eq. \eqref{gen}),
we obtain for particle pair interchange the total wave function  phase shifts $p\pi$
(for $\alpha=\pi$ in the representation given by Eq. \eqref{repr}), as is required
by Laughlin correlations \cite{laughlin1,laughlin2}, without  modeling them by  fictitious vortices.

\subsection*{Definition of an individual particle cyclotron trajectory}
\begin{figure}[tb]
\unitlength 1mm
\begin{center}
\begin{picture}(120,59)
\put(0,0){\resizebox{120mm}{!}{\includegraphics{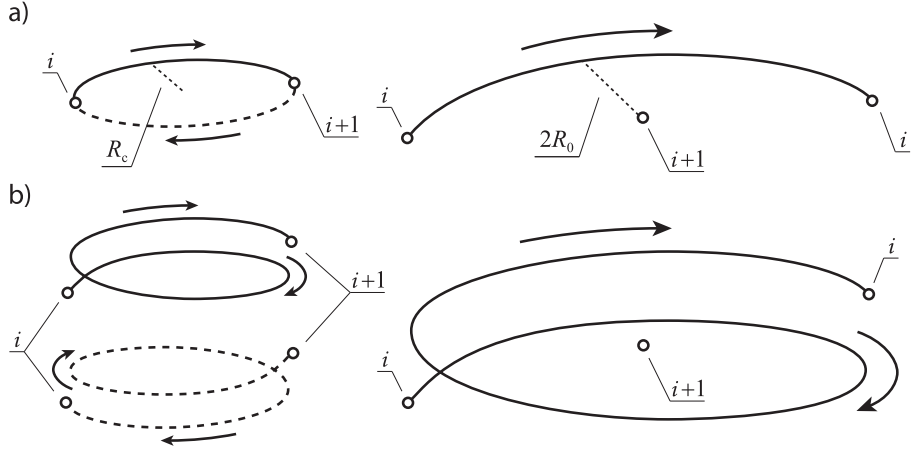}}}
\end{picture}
\end{center}
\caption{\label{fig4} Cyclotron (left) and  corresponding relative (right)  trajectories for interchanges of $i$th and $(i+1)$th 2D-particles
at strong magnetic field, (a) for $\nu =1$, (b) for $\nu =\frac{1}{3}$, respectively  (3D for better visualization); in both cases,
$\nu=1,\frac{1}{3}$, the appropriate cyclotron radius
$R_c$ fits with the inter-particle separation $2R_0=2R_c$, $2R_0$---inter-particle separation  is fixed by the Coulomb repulsion}
\end{figure}

Note, that
each additional loop of a relative trajectory for particle pair interchange
(such a trajectory is needed for definition of  the subgroup generators $b_i^{(p)}$) reproduces  an additional loop in individual cyclotron trajectories
for both interchanging particles---cf. Fig. \ref{fig4}.
In this figure the cyclotron motion of particle pair
is depicted for the interchange of $i$th and $(i+1)$th particles separated by double cyclotron radius $2R_c$, without any additional loops (a)
and with the additional loop (b), respectively. The cyclotron trajectories are repeated in the relative trajectory (right) with a double radius
in comparison to the individual particle trajectories (left).
In quantum language, with regard  to classical multi-loop cyclotron trajectories, one can conclude only on the
number, $\frac{BS}{N}/\frac{hc}{e}$, of flux quanta per  single particle in the system, which for the LL filling $\frac{1}{p}$ is $p$, i.e., the same as
the number of cyclotron loops of each particle. Thus a simple pictorial rule could be here formulated:
 an additional loop on a braid corresponding to particle interchange, introduced in accordance with generators of
 the cyclotron braid subgroup,  results in {\it two} additional flux quanta  piercing the individual particle cyclotron trajectories.
It immediately follows from the definition of the cyclotron trajectory.

 One can define this trajectory as
the individual particle trajectory corresponding to a {\it double} interchange of the particle pair (cf. Fig. \ref{fig5}). In this way, the cyclotron trajectories
of both interchanging particles are closed, similarly as closed the relative trajectory for double interchange is. If the interchange
is simple, i.e., without any additional loops, the corresponding individual  particle cyclotron trajectories are also simple, single-loop (circles on 2D plane).
But when the interchange of particles is multi-loop, as associated with $p$-type cyclotron subgroup ($p>1$), the double interchange
relative trajectory has $2 \frac{p-1}{2}+1$ closed loops and the individual cyclotron trajectories are also
multi-loop, with $p$ loops. It is illustrated in the Fig. \ref{fig5}.

It is worth to emphasize the difference between turns  of windings (e.g., of a wire)  and multi-loop 2D cyclotron trajectories.
The latter ones cannot enhance a piercing total magnetic field flux $BS$ (thus the number of flux quanta per particle coincides with
the number of loops of closed cyclotron individual particle trajectory), while
in the former case, each turn of windings adds a new portion of the flux as a new turn adds a new surface in fact (which is no case in 2D).

\subsection*{Relation of cyclotron braid subgroups with CFs}

We will explain below  that the multi-loop shape of the relative trajectory for interchanges,
as defined by the subgroup generators \eqref{gen}
(and corresponding  multi-loop form of individual particle cyclotron trajectories), is an unavoidable property in the case
when inter-particle separation (resulted from the density $\frac{N}{S}$ and fixed by the Coulomb repulsion) is greater than the
double value of single-loop cyclotron radius. In this case, in particular at $\frac{1}{p}$ LL filling fraction, any
exchanges along simple single-loop cyclotron trajectories are impossible, because the corresponding cyclotron
radius is too short. In order to restore a possibility of
particle interchanges (necessary, on the other hand, for braid structure definition and thus for statistics determination), too short cyclotron radius
must be enhanced. The way to enhance  the effective cyclotron radius, which would again  fit to inter-particle separation, is the
multi-loop character of cyclotron motion and simultaneously resulting  multi-loop braids for particle interchanges (represented by generators
of the cyclotron braid subgroups, Eq. \eqref{gen}).
The additional cyclotron loops take away a part of the external field
flux and thus reduce the effective field which leads to an expected growth of a resulting cyclotron radius.

\begin{figure}[tb]
\unitlength 1mm
\begin{center}
\begin{picture}(120,38)
\put(0,0){\resizebox{120mm}{!}{\includegraphics{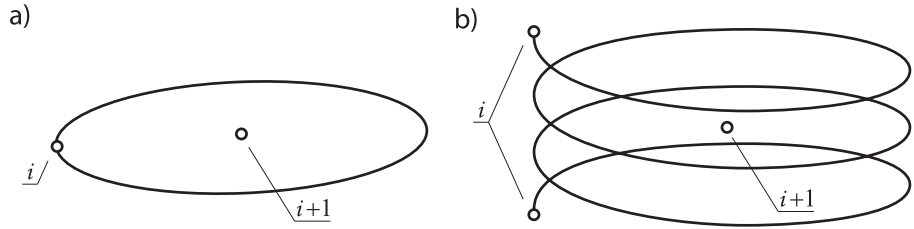}}}
\end{picture}
\end{center}
\caption{\label{fig5} Individual particle closed cyclotron trajectories corresponding to {\it double}
 relative trajectories for interchanges of $i$th and $(i+1)$th 2D-particles
at strong magnetic field, (a) for $\nu =1$, (b) for $\nu =\frac{1}{3}$, respectively  (3D for better visualization);
the number of $B$ field flux quanta per particle is indicated in both cases,
$\nu=1,\frac{1}{3}$;  the resulting cyclotron radius
$R_c$ fits with the inter-particle separation $2R_0=2R_c$ in both cases}
\end{figure}

The total flux of the external field through the surface $S$ is $BS$.
For $p$ type of CFs, if one  considers  the relative trajectory of {\it double} interchange of $i$th and $(i+1)$th particles
(thus closed and with $2\frac{p-1}{2}+1=p$ loops),
 one  gets  the individual particle closed cyclotron trajectories  with the same number $p$ of loops (cf. Fig. \ref{fig5}),
embracing the total flux  $p\frac{hc}{e}$ (each loop takes away a single flux quantum in accordance with the above presented interpretation).
Thus for $p$ type of CFs we deal with  closed $p$-loop cyclotron
 trajectories
 of particles, i.e. $p$ flux quanta per particle,  $BS=Np\frac{hc}{e}$. On the other hand,  the  degeneracy of the
 LL equals to $N_0=\frac{SBe}{hc}$,
 (neglecting spin)
and for  fractional filling $\nu$,  $N_0=\frac{N}{\nu}$.   $\frac{BS}{N}=\frac{hc}{e} \frac{1}{\nu}$  gives $\frac{1}{\nu}$ flux quanta per
particle, which fits with the previous estimation only for $\nu=\frac{1}{p}$.

In the case of $p$-loop trajectory  each loop has its
 size adjusted to the  external magnetic
field flux diminished by  $p-1$ quanta per particle taken away by remaining loops, exactly as in the case
of the Jain's model. Indeed, if $BS=\frac{hc}{e}pN$, then $\frac{hc}{e}=\frac{B}{p}\frac{S}{N}$ and
$\frac{S}{N}$ corresponds to $p$ times lowered field. Following an analogy with  Jain's model, one could argue that  for  $\nu=\frac{1}{2}$ and $p=3$,
two loops per particle take away the total $B$ field flux and the third loop has to be of  infinite  radius (Hall metal \cite{halperin})
 for zero rest-field.

 The additional loops take away flux quanta simultaneously diminishing  the field; this
gives an explanation of the fictitious
Jain fluxes screening the field $B$. Thus the presented cyclotron subgroup implementation of CFs  can be addressed to  Jain's theory with
all advantages of the related conclusions \cite{hon}, in particular of the integer quantum Hall effect in the rest-field, leading to
hierarchy of FQHE \cite{jain,hon}.

One can thus summarize why 2D charged particles must be associated with classical multi-loop braids  for fields corresponding to fractional filling of
 LL. For  $\nu =1$ one has $R_c=R_0$ (where $R_c$  is the cyclotron radius, $ \pi R_c^2 B=\frac{hc}{e}$
and  $2R_0$ is the separation of  particles, adjusted to the density and fixed by the short-range part of the Coulomb repulsion, $\pi R_0^2=\frac{S}{N}$).
 For $\nu<1$ the radius of cyclotron  trajectory
without additional loops $R_c<R_0$,  and then $R_c$ is  {\it too short} for
 particle interchange along these trajectories. Additional loops can however enhance $R_c$ and again allow interchanges,
  since for $p$-loop cyclotron trajectories $\frac{hc}{e}= \pi R_c^2 \frac{B}{p}$, and $R_c$ grows in comparison to single-loop trajectories;
  for $\nu=\frac{1}{p}$, again $R_c=R_0$, though the external field is $p$ times bigger than for $\nu=1$ (at constant $N$).
The fictitious fluxes of Jain's CFs  played actually the similar role
leading to an increase of cyclotron radius in the reduced resultant field. One can conclude thus that for $\nu=1$ the cyclotron trajectories
are single-loop and braids are generated by $b_i^{(p=1)}=\sigma_i$, while for $\nu=\frac{1}{p}$, $p>1$, the cyclotron trajectories must be
multi-loop, simultaneously resulting  in braids generated by $b_i^{(p)}=\sigma_i^p$.

 Note finally
that for a fixed magnetic field orientation the one-side cyclotron rotation is admitted, thus the
cyclotron subgroup should be confined to its semigroup structure only. It does not cause, however,  any perturbations of
relevant 1DURs of cyclotron subgroups, which are  crucial for identification of composite particles.

 The additional loops associated with the appropriate subgroup generators
lead to the phase shifts for particle interchanges, just as for Jain's CFs
 and  permit corresponding Laughlin-type function requirements to be satisfied. These loops replace the fictitious screening fluxes.
Note once more that multi-loop trajectories (similarly as single-loop ones) have only meaning in classical braid terms. Quantum particles
do not traverse any trajectories, also any multi-loop cyclotron trajectories. The corresponding wave functions
transform, however, in an agreement with 1DURs of the brad group or of the  subgroup \cite{imbo1,sud}, resulting
in appropriate statistics behavior.

Let us emphasize that though  CFs actually are not
compositions of particles and vortices, we have not modified the original name 'composite fermions'.
Moreover we use the similar name  'composite anyons' for particles associated with fractional 1DURs (i.e., with fractional  $p\alpha$ in
Eq. \eqref{repr}) of the cyclotron subgroup instead of the full braid group. The phase shift $\theta$ can be calculated as the
Berry's phase along closed trajectory in configuration space for model multi-particle wave function corresponding to low energy
excitations above the ground state at fractional filling of LL. These excitations---quasiparticles/quasiholes were traditionally associated with anyons
in the case of a fractional Berry's phase. It is, however, clear that it is impossible to  distinguish between
fractional $\theta$ and $p\alpha$---both these phase shifts can be the same fraction. As considered
quasiparticles/quasiholes are excitations at the magnetic field presence, thus these states should be rather associated with cyclotron braids 1DURs,
and therefore are composite anyons
and not ordinary anyons, as previously regarded.
This change, anyons for composite anyons, would result in  convenient for QIP more dense relevant MDURs corresponding to braid subgroup instead of the full
braid group.

  \subsection*{A role of the short-range part of the Coulomb interaction}

 The crucial  character of the short-range part of the Coulomb interaction
     for Laughlin correlations is visible from the fact that the Laughlin function is an accurate ground state wave function at $\frac{1}{p}$ LL filling,
     if  to
     confine the Coulomb interaction  represented by the so-called Haldane's pseudopotential \cite{haldane,prange},
     $V=\sum_{i>j}{\sum_m^{\infty}{ V_m P^{ij}_m}}$, ($P^{ij}_m$ is the projector
     on the states of $i$th and $j$th particles with relative angular momentum $m$), to the components $V_m$, with $m=1,...,p-2$ only. These $V_m$ terms,
     the Coulomb interaction energy of an
     particle pair with  relative angular momentum $m$, contribute the short-range part of the
     interaction of electrons, and the remaining terms---long-range interaction tail, corresponding to greater particle separation, i.e., with $m=p,...$, do not influence
     strongly the Laughlin function \cite{prange,haldane,mo}. The Laughlin correlations are associated with the incompressible states which
     correspond to discrete spectrum of Coulomb interaction projected on LL states, i.e., interaction expressed in terms of Haldane's pseudopotential
     with components assigned by relative angular momentum of particle pairs. This property, essential for FQHE, was even
     named by Laughlin as "a quantization of particle separation" \cite{laughlin2,prange}. Quantization of the Coulomb interaction after
     projection on relative angular momentum of particle pairs in LL Hilbert  subspace results
     in incompressible FQHE states numbered by integers (eigen-values of relative angular momentum of particle pairs), the same which occur in the Laughlin functions
     (the exponent in the Jastrow polynomial).
      It is important to note that according to an attitude to FQHE
     using Haldane's pseudopotential (confined to the short-range part of the Coulomb repulsion), the Laughlin correlations revealed in
     the multi-particle wave function are  unambiguous
     possibility for accurate ground state at fractional LL filling, not only a variational result of the ground state modeling \cite{prange,haldane}.
     It supports an idea that
     Laughlin correlations are a fundamental topology-originated property of interacting charged 2D particles. One can thus expect that this
     Landau quantization behavior of interacting 2D charged system must also manifest itself within braid group quantization approach to the same system, via
     the introduced  cyclotron subgroup structure.

Since the Laughlin correlations can be expressed within CF approach, thus the Coulomb repulsion (the short range part of Haldane's pseudopotential)
is of a  fundamental significance
also for the CF construction. It should be, however, emphasized that the
Coulomb interaction with the discrete  spectrum, i.e., with separation by energy gaps the distinct relative angular momenta of particle pairs
 for sufficiently high
magnetic field (noticeable via projection of the interaction on fractional filled LL as in the definition of Haldane's pseudopotential) does not play
a role of standard dressing of particles with interaction, typical for quasiparticles in solids, just because the interaction has not a continuous spectrum in this
projection.

An effective description of a local gauge field attached to particles
 is supplied by the
    Chern-Simons (Ch-S) field theory (chiral field, i.e., breaking time reversion and parity).
    This approach revived \cite{cs,lopez} in the area of FQHE successfully describing particles with fluxes, in particular
      anyons and Jain's CFs \cite{hon}. It still, however, does not explain, what
    the spontaneously arising  fluxes are.

   It was demonstrated \cite{prange,mo} that the short-range part of the Coulomb interaction stabilizes CFs against action of
    Ch-S field (its antihermitean term), which  mixes states with distinct angular momenta  within LL \cite{mo} in disagreement with the Jain's CF model in
    CH-S field approach \cite{hon,mo}.
    The Coulomb  interaction   removes the degeneracy of these states and results in energy gaps which  stabilize CF picture,
     especially effectively within the lowest LL. For higher LLs  CFs are not so useful due to possible
     mixing between   LLs induced by interaction \cite{id}.

The short-range part of the Coulomb interaction stabilizes CFs also in our description, similarly
as it removes instability  caused by Ch-S field
for angular momentum orbits in LL \cite{mo}.  Indeed, if
the short-range part of the Coulomb repulsion was reduced, the separation of particles would not be rigidly kept (adjusted   to a density only in average) and then
another cyclotron trajectories, additional to  those  for fixed particle separation (multi-loop at $\nu=\frac{1}{p}$), would
 be admitted, which violates the subgroup construction.

  Thus the short-range part of the Coulomb interaction turns out to be crucial for CF formation in any description. Confining of the full braid group
  to the subgroup  with multi-loop structure of cyclotron motion is justified only for
 particle separation adjusted  to the double cyclotron radius. It is a role of the short-range part of the Coulomb repulsion which
  does not allow closer inter-particle separation than that which follows form the density. In this manner the short-range  part of the
   Coulomb interaction is involved
  in the construction of the cyclotron braid structure. The  long-range tail of the Coulomb interaction is left as a residual interaction
  of particles, which agrees with the Jain's model of weakly interacting CFs \cite{jain,hon}.

\section{Conclusions}

We have developed the braid group description for the case of $N$ charged 2D particle system at strong magnetic field presence, via definition
of the cyclotron braid subgroups. This formalism allowed for interpretation of the Laughlin correlations
of 2D charged systems within the braid group approach to $N$ particle quantum systems.
In this manner we formulated a new implementation of CFs  employing  braid group methods.
Braid description  of CFs was not previously established because of  $2 \pi$ period of 1DURs.
      In the present paper  we have  avoided this problem via
             reduction of 1DURs to specially chosen braid subgroups selected in accordance with a 2D cyclotron motion. These cyclotron braid
            subgroups, generated by the new generators $b_i^{(p)}=\sigma_i^p$
             ($p=1,3,5,...$ enumerates a sort of composite anyons, $\sigma_i$ are generators of the full braid group),
             are  separated braid objects which allow for distinguishing in statistics
             of CFs (with $p>1$) from ordinary fermions.
               It supports an idea that CFs are rightful 2D quantum
             particles which cannot be mixed with ordinary fermions,  or with other sorts of CFs (though all correspond to antisymmetric functions).
             Distinguishing of CFs from  fermions  is important in particular for numerical diagonalization of interaction of
CFs (not all antisymmetric functions can be admitted in diagonalization procedure, but only those which have the same phase shift due to particle interchanges, unless the mixing of
various sorts of CFs took place [this is prohibited, similarly to the mixing of fermions and bosons in 3D]).

             CFs turn out thus to be real 2D particles and not quasiparticles, i.e., they are not fermions dressed with interaction only,
             but are arranged as separate particles in topological terms.
             Identification of the special braid group object, the subgroup of the full braid group, associated with  CFs,
             resolves also the problem of fictitious magnetic flux quanta, vortices, attached to these particles within the standard Jain's model.
             The Aharanov-Bohm phase shifts
             caused by  hypothetical fluxes  are replaced  with the phase shifts due to additional $\frac{p-1}{2}$
             loops during interchanges of particles (described in classical braid terms). These loops are an unavoidable
              property of interchanges of uniformly distributed (due to the Coulomb repulsion)
             2D particles  in  a strong external magnetic field when ordinary cylotron radius is too short for particle interchanges
              (each particle
             traverses, in a classical braid picture, a closed $p$-loop cyclotron trajectory or in quantum language, it takes  away
             $p$   quanta of the $B$ field flux; $p-1$ of them
            play the equivalent role as  $p-1$ screening flux quanta attached to each CF in  Jain's model).

              The 1DURs, $b_i^{(p)}\rightarrow e^{ip\alpha}$,
             $\alpha\in(-\pi,\pi]$, of the cyclotron braid subgroups generated by $b_i^{(p)}$
             ($p$---odd integer) supply, more generally,  an implementation of  composite anyons, including CFs of   rank $p$, for $\alpha=\pi$.
             In particular,  CFs (for $\alpha=\pi$) gain  the phase
             shift $p\pi$ (due to the additional loops) the same as required by  Laughlin-type correlations.
         The composite particles within the presented implementation
              are thus not  connected with the full braid group but with their cyclotron subgroups.
             It makes CFs described rightfully with other types of 2D quantum particles  within the uniform braid group approach,
              despite the $2\pi$ period limitation for 1DURs.

             An important role  of the short-range part of the Coulomb  interaction is indicated. This interaction fixes the inter-particle separation,
             (only in average determined by the planar density),
             which allows for definition of multi-loop cyclotron braid trajectories for particle interchanges
             in the case when single-loop cyclotron radius is too short in comparison to  inter-particle separation,
             precluding particle exchanges along single-loop trajectories, as for $\frac{1}{p}$ LL filling.
             The additional loops reduce the  total magnetic field flux and enhance the effective cyclotron radius, restoring
             possibility of particle interchanges.  Thus multiloop trajectories are unavoidable property of cyclotron braids
              leading, in a natural way, to the Laughlin correlations,  without  artificial constructions with vortices.

             On the other hand, the cyclotron subgroups may have richer unitary representations, including MDURs, in comparison to the full braid group, which
             would result in more dense MDURs corresponding to composite non-Abelian anyons for possible QIP applications.

 Supported by the Polish KBN Project No N202 071 32/1513
 and Network LFPPI

\bibliographystyle{my-h-elsevier}

\begin{thebibliography}{20}





\bibitem{spanier}
E. Spanier,
Algebraic Topology,
Springer Verlag, Berlin 1966.

\bibitem{birman}
J. S. Birman,
Braids, Links, and Mapping Class Groups,
Annals of Math. Stud. {\bf 82}, Princeton UP, Princeton 1974.

\bibitem{mermin}
N. D. Mermin,
Rev. Mod. Phys. {\bf 51} (1979) 591.

\bibitem{tsui}
D. C. Tsui, H. L. St\"ormer and A. C. Gossard,
Phys. Rev. Lett. {\bf 48} (1982) 1559.

\bibitem{pin}
S. Das Sarma and A. Pinczuk,
Perspectives in quantum Hall effects: novel quantum liquids in low-dimensional semiconductor structures,
Wiley, New York 1997.

\bibitem{prange}
R. E. Prange, and S. M. Girvin,
The Quantum Hall Effect,
Springer Verlag, New York 1990.

\bibitem{laughlin1}
R. B. Laughlin,
Phys. Rev. B {\bf 27} (1983) 3383.

\bibitem{laughlin2}
R. B. Laughlin,
Phys. Rev. Lett. {\bf 50} (1983) 1395.

\bibitem{wilczek}
F. Wilczek,
Fractional Statistics and Anyon Superconductivity,
World Sc., Singapore 1990.

\bibitem{wu}
Y. S. Wu,  Phys. Rev. Lett. {\bf 52} (1984) 2103.


\bibitem{artin}
E. Artin,
Annals of Math. {\bf 48} (1947) 101.

\bibitem{jac1}
L. Jacak, P. Sitko, K. Wieczorek and A. Wojs,
Quantum Hall Systems: Braid groups, composite fermions, and fractional charge,
Oxford UP, Oxford 2003.

\bibitem{imbo}
T. D. Imbo, C. S. Imbo and E. C. G. Sudarshan,
Phys. Lett. B {\bf 234} (1990) 103.

\bibitem{enar}
T. Einarsson,
Phys. Rev. Lett. {\bf 64} (1990) 1995.

\bibitem{dassar}
C. Nayak, S. H. Simon, A. Stern, M. Freedman, and S. Das Sarma,
Rev. Mod. Phys. {\bf 80} (2008) 1083.

\bibitem{jain}
J. K. Jain,
Phys. Rev. Lett. {\bf 63} (1989) 199.

\bibitem{hon}
O. Heinonen,
Composite Fermions,
World Sc., Singapore 1998.

\bibitem{halperin}
B. I. Halperin, P. A. Lee and N. Read,
Phys. Rev. B {\bf 47} (1993) 7312.

\bibitem{imbo1}
T. D. Imbo and E. C. G. Sudarshan,
Phys. Rev. Lett. {\bf 60} (1988) 481.

\bibitem{sud}
E. C. G. Sudarshan, T. D. Imbo and T. R. Govindarajan,
Phys. Lett. B {\bf 213} (1988) 471.

\bibitem{kitaev}
A. Kitaev,
Russian Math. Survey, {\bf 52:61} (1997) 1191;
A. Kitaev,
Annals Phys. {\bf 303} (2003) 2;
M. Freedman, A.Kitaev, M. Larsen and Z. Wang,
Bull. Amer. Math. Soc. (N.S.) {\bf 40} (2003) 31.

\bibitem{nielsen}
M. A. Nielsen and I. L. Chuang,
Quantum Computation and Quantum Information,
Cambridge UP, Cambridge 2000.

\bibitem{xia}
J. S. Xia {\it et al.},
Phys. Rev. Lett. {\bf 93} (2004) 176809;
E. H. Rezayi and N. Read,
arXiv:cond-mat/0608346v1 (2006).

\bibitem{haldane}
F. D. M. Haldane,
Phys. Rev. Lett. {\bf 51} (1983) 605.

\bibitem{mo}
T. Morinari,
Phys. Rev. B {\bf 62} (2000) 15903.

\bibitem{cs}
A. L. Fetter, C. B. Hanna and R. B. Laughlin,
Phys. Rev. B {\bf 39} (1989) 9679.

\bibitem{lopez}
A. Lopez and E. Fradkin,
Phys. Rev. B {\bf 44} (1991) 5246.

\bibitem{id}
T. Sbeouelji and N. Meskini,
Phys. Rev. B {\bf 64} (2001) 193305.



\end{thebibliography}

\end{document}